\DocumentMetadata{}
\documentclass[manuscript,screen,acmsmall]{acmart}
\usepackage{hyperref}
\usepackage{hyperxmp}
\AtBeginDocument{%
  }

\setcopyright{acmlicensed}
\copyrightyear{2025}
\acmYear{2025}
\acmDOI{XXXXXXX.XXXXXXX}
\acmConference[CSCW'26]{}{}{}

\begin{document}

\title{From the CDC to emerging infectious disease publics: The long-now of polarizing and complex health crises}

\author{Tawfiq Ammari}
\affiliation{%
  \institution{Rutgers University}
  \city{New Brunswick, NJ}
  \country{USA}}
\email{tawfiq.ammari@rutgers.edu}

\author{Anna Gutowska}
\affiliation{%
  \institution{Rutgers University}
  \city{New Brunswick, NJ}
  \country{USA}}
\email{anna.gutowska@rutgers.edu}

\author{Jacob Ziff}
\affiliation{%
  \institution{Rutgers University}
  \city{New Brunswick, NJ}
  \country{USA}}
\email{jacob.ziff@rutgers.edu}

\author{Casey Randazzo}
\affiliation{%
  \institution{Rutgers University}
  \city{New Brunswick, NJ}
  \country{US}}
\email{cer124@rutgers.edu}

\author{Harihan Subramonyam}
\affiliation{%
  \institution{Stanford University}
  \city{Stanford, CA}
  \country{USA}}
\email{harihars@stanford.edu}

\renewcommand{\shortauthors}{Ammari et al.}

\begin{abstract}
This study examines how public discourse around COVID-19 unfolded on Twitter through the lens of crisis communication and digital publics. Analyzing over 275,000 tweets involving the CDC, we identify 16 distinct discourse clusters shaped by framing, sentiment, credibility, and network dynamics. We find that CDC messaging became a flashpoint for affective and ideological polarization, with users aligning along competing frames of science vs. freedom, and public health vs. political overreach. Most clusters formed echo chambers, while a few enabled cross-cutting dialogue. Publics emerged not only around ideology but also around topical and emotional stakes, reflecting shifting concerns across different stages of the pandemic. While marginalized communities raised consistent equity concerns, these narratives struggled to reshape broader discourse. Our findings highlight the importance of long-term, adaptive engagement with diverse publics and propose design interventions—such as multi-agent AI assistants—to support more inclusive communication throughout extended public health crises.
\end{abstract}




\received{24 March 2025}
\received[revised]{XX XX XXXX}
\received[accepted]{XX XX XXXX}

\maketitle

\section{Introduction}
\begin{quote}
``But for politics and bad beer, the matter would never have been heard of''
\\Walter Kempster, Milwaukee Health Commissioner (1841-1918)
\end{quote}

The use of technology to disseminate public health messages has historically shaped public health outcomes. In the quote above, the \textit{matter} Kempster describes is the breakdown of civic life in Milwaukee, WI in 1894 due to a smallpox outbreak that got out of control \cite{leavitt2003public}[P.187-188]. While Kempster was a capable physician, he overlooked the importance of \textit{communication} in explaining health risks to the city population and opening communication channels to receive their feedback. A widely different experience describes the effective control of a smallpox outbreak in 1947 in New York City. In that iteration, government organizations used frequent, clear, and simple messages to keep citizens updated through different channels (e.g., newspapers, radio shows, and even vaccination lapels): ``Be safe. Be sure. Get vaccinated.'' These clear communication techniques were coupled with collaboration with volunteer organizations (e.g., the Red Cross) which provided more channels to different communities in the city \cite{leavitt2003public}[P.189]. The communication of public health messages is still a challenge for organizations today. For example, the recent measles outbreak in the U.S. suggests the continued struggle to manage vaccine disinformation, inequitable health care access, and limits to health literacy. These cases raise important questions for human-computer interaction (HCI) scholars on the role of design in organizations' use of social media to disseminate public health messages and interact with the public. 

To investigate these challenges, this study analyzed the discourse of the Centers for Disease Control and Prevention (CDC) during the COVID-19 Pandemic. Taking a longitudinal approach, we examined how the CDC sent and received communication messages across the first two years of the pandemic on Twitter.\footnote{Twitter re-branded to X after the collection of the data used in this study and so, we refer to it as Twitter throughout this paper.} This exchange of communication messages on Twitter is an example of the Habermasian public sphere -  a space for ``the coming together of peoples and the discussion of ideas...often related to governance and Democratic ideals.'' \cite{poor2005mechanisms}  These publics collapse around ``issues'' of discussion (e.g., \#COVID) \cite{dahlberg2001computer}, or arenas of discussion (e.g., online neighborhood group \cite{randazzo2025werelosingneighborhoodswere} or public health institutional account) \cite{moe2023operationalizing}. Therefore, we refer to the area of discourse we analyze as the \textit{\textbf{CDC public}}.

Twitter is a virtual ``public square'' \cite{jenzen2021symbol} where different ad hoc ``publics'' can be identified using hashtags \cite{bruns_structural_2014} which are consumed by interested audiences \cite{livingstone2005relation}. They can also be formed around mentions that are searchable by other users \cite{kountouri2023polarizing}. Retweets, quote-tweets, and responses to Tweets are associated with propagating messages in different ways. 

Crisis management includes official policies and communication received from government entities and major health organizations.  Over recent years, there has been a significant increase in government use of social media as a vehicle for disseminating public health crisis information. In turn, there is an increased expectation among the public for this greater level of communication \cite{loni_et_al_18}. Using social media for crisis management information can be of great use to the public. For instance, during the 2009 H1N1 influenza outbreak, the Alexandria Virginia Health Department informed citizens of where to receive vaccinations using social media \cite{raina_et_al_11}.

Crisis communication is the process of providing facts to the public about an unexpected emergency, beyond an organization's control, that involves the organization and requires an immediate response \cite{zahry2023risk}. During health crises, information becomes a `necessity, not a luxury'. Public health communication can be compared to political persuasion by framing their messages to push a specific goal \cite{dennison2019rising,entman1993framing}. In turn, audiences have `interpretive schemata' that allow them to understand the message. Messaging can be honed by stressing the salience of specific frames \cite{oppermann2011issue}. These frames can become polarized by different stakeholders in the public \cite{leavitt2003public}. Discursive polarization is affective (e.g., social identity is polarized by framing discourse as ``us vs. them'') and ideologically polarized by stressing the disagreement about the definition of the problem at hand (e.g., we are facing a pandemic) and providing evaluative statements (e.g., \#maskssavelives) \cite{bruggemann2023debates}. Given the centrality of framing to making sense of a public space, we ask:

\textbf{RQ1: What are the major frames of discussion in different publics and how are they polarized?}

Boyd defines networked publics as ``publics that are restructured by networked technologies; they are simultaneously a space and a collection of people.'' \cite{boyd2010social} Those societal stakeholders are ``actors and organizations viewed as having a vested interest in the decision output.'' \cite{crow2020evaluating}[P.2] Hallahan \cite{hallahan2000inactive} breaks stakeholders into four main groups: (1) active stakeholders: have deep knowledge and involvement in the policy process; (2) aware stakeholders: knowledgeable stakeholders who might not be directly affected by policy change (e.g., policy experts and opinion leaders); (3) aroused stakeholders: while they have a stake in policy, they have low levels of knowledge about it; and (4) inactive stakeholders: lacking both knowledge and involvement \cite{park2020stakeholder}[P.642]. Given the importance of identifying different archetypes of stakeholders in a potentially polarized public, we ask:

\textbf{RQ2: Who are the main stakeholders in our public? How are they polarized around different issues? }

These disparities are not only shaped by structural inequality but also shift over time as communities move through different stages of a crisis. The evolving nature of pandemics—marked by changing levels of risk, policy, and public discourse—means that the needs of marginalized communities are not fixed. For example, initial concerns may center on access to protective equipment or credible information, while later stages may emphasize recovery support, mental health resources, or rebuilding trust in institutions. As Fox argues, the “rolling uncertainty” of emerging infectious diseases disrupts any stable ground for action or expectation \cite{fox2012medical}, while Dalrymple cautions that mismatches between institutional narratives and lived realities can compound distrust over time \cite{dalrymple2016facts}. Recognizing how marginalized communities’ concerns develop temporally is essential for shaping more responsive and equitable crisis engagement.

\textbf{RQ3: How do the needs, concerns, and responses of marginalized communities evolve over the course of a pandemic, and what implications does this have for designing inclusive and adaptive public health communication strategies?}

\begin{figure}
    \centering
    \includegraphics[width=0.9\linewidth]{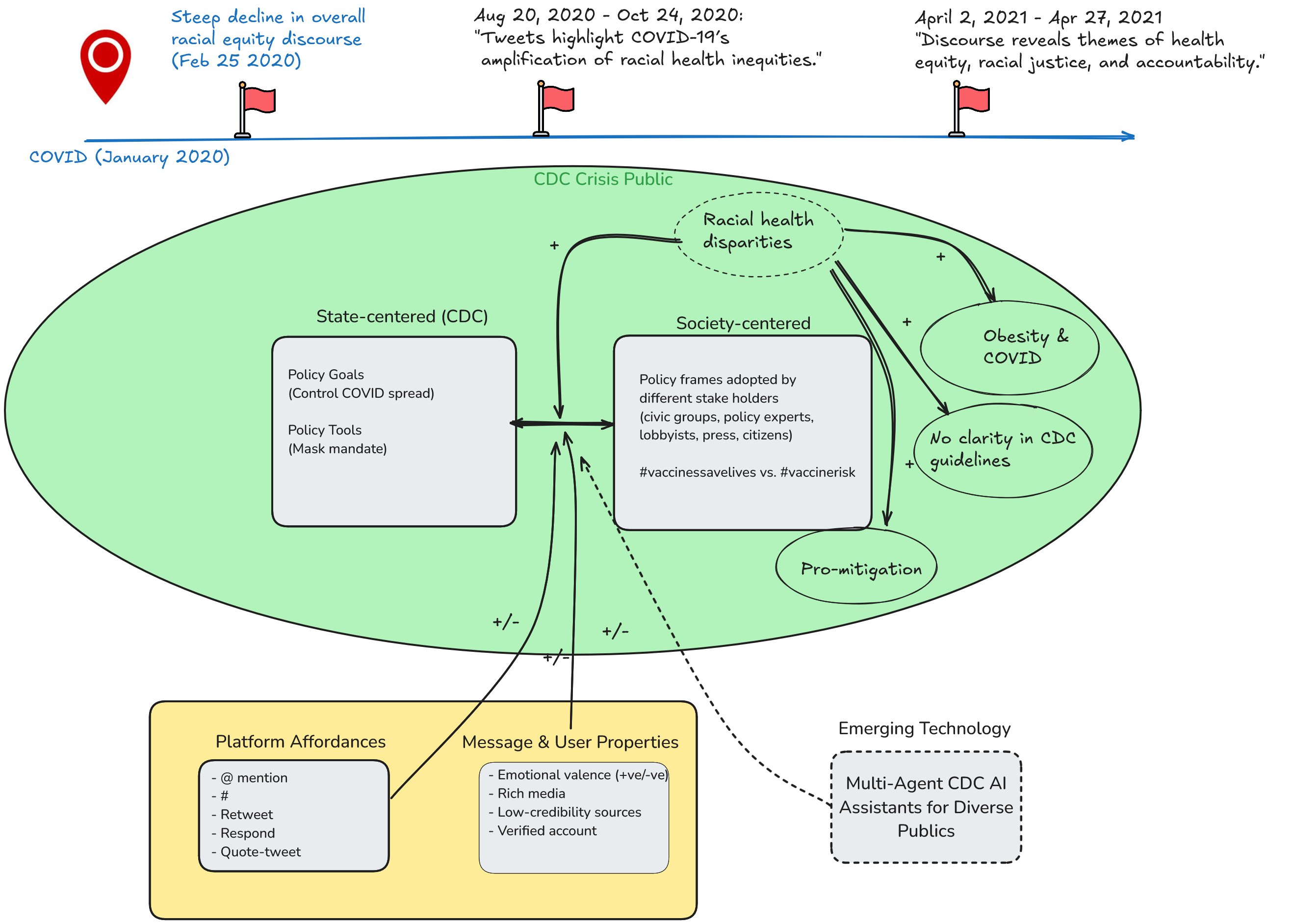}
    \caption{This image shows public policy discourse framing analytical framework adapted from \cite{park2020stakeholder}. We introduced new elements (in yellow) to the model based on our findings. We show how platform affordances, message and user properties, and platform governance are important factors to consider when communicating public health policy for emerging infectious diseases like COVID. Additionally, we discuss how emerging technologies (in gray), especially generative AI introduce both challenges and design opportunities for crisis organizations like the CDC. Finally, we discuss the effects of temporal crisis distance affects CDC messages over time, especially around flashpoints associated with major changes (e.g., new virus variants). The +/- signs represent how each of these components can make crisis communication more or less effective.}
    \label{fig:park_lee}
\end{figure}

We adapt the public policy discourse framing analytical framework described in Park and Lee \cite{park2020stakeholder}[P.641] to model how Twitter is used as a communication channel to and from the CDC throughout a two-year public health emergency - COVID (see Figure \ref{fig:park_lee}). The framework suggests that public policy comprises communication between the state and society (public \& private). In this framework, ``communicative interaction among political actors operates in a multilayered policy landscape...[to advocate] for ''policy goals (in our case, COVID mitigation policy) and formulating the policy tools (e.g., mask and vaccine mandates) needed to achieve these goals.

We incorporate into the model three new components: (1) the affordances of propagation in digital platforms (e.g., retweets, mentions etc.); (2) message and user properties including the sentimental valence of the message, the credibility of sources, and whether the user is verified. We used these features to cluster our users into clusters representing different publics. Finally, we present a new third component representing emerging technologies, specifically, the CDC AI assistant for diverse publics presented in the discussion section (see \S\ref{sec:engage_gen_ai}) that can assist crisis organizations like the CDC in maintaining a line of communication with different publics. Additionally, we show how while most publics acted as echo chambers, one marginal issue public (Cluster 1) connected with the CDC and other publics. However, the model also displays the limits of this engagement by Cluster 1 as while the members of the cluster were focused on the needs of their members (two later changepoints), they could not change the trend of declining interest in racial health disparities throughout the Twitter publics (see \S\ref{subsec:marginalized_pub}). 

\section{Related Work}
In this section, we review prior research that informs our study. We begin with work on how digital publics form and operate during crises. Next, we examine how framing and polarization influence discourse, especially in the context of public health emergencies. Finally, we consider studies on digital communication strategies in crisis response, focusing on trust, misinformation, and public engagement.

\subsection{Digital Publics and Crisis Management}
When opinions and knowledge are shared in a communicative space, ``an arena in which citizens discuss matters of common concern'' \cite{wessler2019habermas}, it is referred to as a Public Space \cite{habermas1989jurgen}. A public can be defined as a space to access information along with an amorphous network of individuals \cite{van1985semantic}. Publics allow for knowledge to be widely shared, making ``collective thoughts and action possible.' \cite[P.14]{van1985semantic}. When people have a shared awareness of knowledge, they have a better sense of collective power, while the lack of access to shared knowledge leads to detachment \cite[P.14]{van1985semantic}.

In these public spheres, stakeholders come together to talk about \textit{issues} (c.f., \cite{dewey1927half,rauchfleisch2021public}) defined as ``socially constructed matters of public interest around which there is contention'' \cite[P.2009]{stoltenberg2024spaces}. These publics evolve over time around specific topics of interest signified by hashtags that convey the content of the messages \cite{bruns2011use}. 

This in essence leads to the creation of two types of publics: (1) active publics which are connected to the organization managing a crisis (e.g., local and State agencies following FEMA or State public health institutions following the CDC) \cite{kinsky2021crisis}; and (2) latent publics: which are formed by people who, while affected by a crisis, might not have control over the response  \cite[P.445]{dalrymple2016facts} but still engage in dialogue about the crisis in real-time \cite{bortree2009dialogic, seltzer2007dialogic, taylor2005diffusion,kim2011problem}.

\subsubsection{Hashtags, mentions, and publics}
Hashtags like \#BLM also signify affiliation with wider communities that signify their values \cite{zappavigna2018communing,freelon2016beyond}. These hashtags serve as focal points, allowing people to share their feelings through storytelling in what Papacharissi refers to as ``affective publics'' \cite{papacharissi2016affective}. During crises, people search for hashtags to find ``situational updates'' that are most important to their crisis \cite{lachlan2016social}. Organizations managing a crisis (e.g., FEMA or the CDC), can create or co-opt hashtags that are followed by a large number of users \cite{dalrymple2016facts}. 

Using mentions on Twitter is another way of creating digital publics that works by targeting central members of a network \cite{kountouri2023polarizing}. For example, when analyzing the use of mentions on Twitter, addressing specific users with negative or harsh messages can increase the polarization of publics \cite{kountouri2023polarizing}. This affective polarization increases dislike towards the out-group (e.g., Republicans attacking Democrats; \cite{druckman2019we,iyengar2012affect,iyengar2019origins})\cite{bruggemann2023debates}. 

\subsection{Framing in Crisis Publics} \label{subsec:frame_public}
Affective polarization also leads to polarized framing of messages on social media \cite{entman1993framing}. The reason framing can produce affect in those receiving a message is that the sender (e.g., politicians) ``attempt to mobilize [people] by encouraging them to think about...policies along particular lines [like] highlighting certain features of the policy[, specifically,] its relationship to'' values that are important to them (\cite{jacoby2000issue} cited in \cite[P.106]{chong2007framing}). In other words, framing allows one to define the problem. Hashtags can be a shortcut to describing evaluative statements on potentially polarizing issues like masking \cite{lang2021maskon} If different people have different worldviews, if for example, one group considers public health mitigation mandates as a central public health policy, while others view it as a question of personal freedom, they will be "talking past each other" \cite{bruggemann2023debates,giglietto2017hashtag}, therefore, crisis organizations like the CDC need to craft different messages that satisfy various publics \cite{Sellnow2009}.

Earlier work using sentiment analysis evaluates how different publics react to public policies like mitigation measures (e.g., masking, vaccination and school closures; \cite{ewing2021navigating,kwok2021tweet,manguri2020twitter}) or economic support to offset lockdown economic losses \cite{HUANG2022102967}. Indeed, public health interventions have been politically polarizing \cite{haupt2021characterizing,valiavska2022politics}, especially on social media sites like Twitter \cite{valiavska2022politics}, where ``Liberate'' movement and other groups countered public health measures (e.g. lock-downs) by sharing hashtags like \#Liberate. These groups were especially championed by political leaders such as President Donald Trump at the time \cite{haupt2021characterizing}. Earlier work shows that the more a message satisfies different affective publics, the more accepted the response message \cite{jin2010making,utz2013crisis,kim2011emotions}. 

\subsubsection{Misinformation and Epistemic (Dis)trust}
Epistemic trust refers to the confidence placed in knowledge produced by scientists—rooted both in their technical expertise and their perceived honesty and impartiality as conveyors of scientific information \cite{hendriks2016trust,wilholt2013epistemic}. However, this trust has been eroding in recent years, a trend that was further exacerbated by the urgency for real-time scientific guidance during the COVID-19 pandemic \cite{giorgi2025contestation}. Erosion of trust in formal sources of crisis-related information has fueled the proliferation of conspiracy theories. Vulnerable groups, in particular, often exhibit heightened skepticism toward public health institutions and healthcare systems \cite{abelson2009does}. For instance, HIV/AIDS-related conspiracy theories propagated the belief that the epidemic was deliberately orchestrated by the government to harm specific populations \cite{kou_at_al_17}. Such narratives can have serious ramifications; in this case, mistrust led some AIDS patients to avoid critical testing and treatment services \cite{christina_2010}. Similarly, during the Zika virus outbreak, social media emerged not only as a platform for up-to-date information but also as a breeding ground for conspiracy theories surrounding the virus \cite{kou_at_al_17}. Kou et al. argue that these theories represented an effort by the public to interpret and cope with the disruptive societal changes brought on by the outbreak \cite{kou_at_al_17}. This sense-making process can quickly spiral into misinformation when health authorities themselves are grappling with incomplete knowledge about the emerging threat \cite{gui_et_al_17}. These dynamics were again evident in the discourse surrounding COVID-19 mitigation efforts \cite{liao_23,Vishwamitra_24,javed_et_al_23,wang_et_al_23}.

\subsection{Digital Communication in Public Health Crises}
Gui et al. advocate for involving the public in crisis communication decision-making to reduce distrust between organizations like the CDC and the communities they serve \cite{gui_et_al_18}. A notable example of this participatory approach is a COVID-19 hackathon in Germany, where participants co-developed solutions across various social domains \cite{haesler_2021}. However, public health agencies more commonly use social media as a one-way broadcast tool rather than a space for reciprocal engagement \cite{Graham_14}. Reframing social media as a two-way communication channel—by actively incorporating public sentiment and feedback—can help identify information gaps and overlooked concerns early, fostering greater transparency and trust.

The Crisis and Emergency Risk Communication (CERC) model emphasizes that effectively managing uncertainty during a crisis requires two-way communication between organizations and the public \cite{reynolds2005crisis, renn2015stakeholder}. This approach allows the public to share feedback on crisis response efforts, fostering more responsive and transparent communication. Public-facing agencies such as the CDC and FEMA, in particular, are encouraged to adopt what Dozier et al. \cite{waters2011tweet} refer to as symmetrical two-way communication—a dialogic exchange that promotes mutual understanding and enables organizations to adapt their messaging strategies accordingly \cite{grunig2008excellence, waters2011tweet}. Despite these recommendations, prior research shows that CDC’s online communication has largely remained top-down, offering limited opportunities for public input or engagement \cite{gesser2014risk}. Additionally, the CERC model does not provide adequate guidance for any public health crisis that might continue for months --- much like COVID \cite[P.5]{miller2021being}. 

This challenge is amplified in the context of emerging infectious diseases (EIDs) like  COVID-19, where medical expertise and adherence to established protocols can project an “illusion of certainty” \cite{dalrymple2016facts}. For instance, when CDC protocols failed to prevent the spread of Ebola, the agency responded by reinforcing existing knowledge about the disease rather than addressing the evolving nature of the threat \cite[P.455]{dalrymple2016facts}. Research shows that projecting certainty amid chaotic and uncertain conditions can backfire, intensifying public doubt and eroding trust in institutions like the CDC \cite{dalrymple2016facts}. Pivotal events—such as the first case of disease transmission to a healthcare worker—serve as “flashpoints of change” that necessitate immediate and transparent shifts in communication strategy \cite{vanderford2007emergency, dalrymple2016facts}.

To address these challenges, the \textit{Roadmap for Public Empowerment Policies in Crisis Management} calls for adaptive and inclusive communication throughout all crisis phases—preparedness, response, and recovery. It underscores the importance of building trust prior to crises, delivering culturally sensitive and accessible messaging during crises, and actively involving communities—especially those that are marginalized or vulnerable—in recovery processes \cite{vos2014roadmap, wieland2022community}. Key strategies include tailoring messages to specific audiences, partnering with trusted local intermediaries, and co-producing response plans with affected communities \cite{rowel2012introduction}.

\section{Dataset}
We used the Python library Twarc to mine tweets via the Twitter API, authenticated through a personal bearer token granted by Twitter. Our dataset includes tweets that met the following criteria: authored in English, originating from U.S.-based users, posted between January 2020 and January 2022, and not classified as retweets. We focused on tweets directly engaging with the Centers for Disease Control and Prevention (CDC) by including messages that were: (1) sent to @CDCgov, (2) posted by @CDCgov, (3) used the hashtag \#CDC, (4) tagged @CDCgov, or (5) contained the string “CDC” in the tweet text. This resulted in the collection of 21,843 tweets sent to @CDCgov, 11,810 from @CDCgov, 20,012 with the hashtag \#CDC, 53,104 tagging @CDCgov, and 162,295 mentioning “CDC.”

Additionally, we included any tweets sharing the same Conversation ID, indicating participation in the same discussion thread. This yielded a final dataset of 275,124 unique tweets. Although full retweet content was not collected, we retained retweet counts, totaling 3,331,350 retweets across the dataset.
\section{Methods}
\label{sec:meth}
In this section, we describe the mixed-methods approach used to analyze CDC-related discourse on Twitter. We detail our dataset construction, feature extraction, and topic modeling techniques, followed by sentiment and credibility analysis. We then outline our clustering method to identify discourse communities, our interpretive strategies for qualitative analysis, and the temporal techniques used to detect shifts in public attention over time.

\subsection{Features used in data analysis} \label{subsec:meth_features}
In this section, we describe the features used to cluster distinct publics and to examine how discourse evolved both within these clusters and across the broader conversation over time. We begin with an overview of the topics and themes generated through topic modeling, followed by our approach to measuring sentiment and identifying low-credibility sources. We then outline additional platform signals (e.g., retweets) and user attributes (e.g., verification status) incorporated into the analysis.

\subsubsection{Topic Modeling with BERTopic (N=71)} \label{sec:topics}
We used the Bidirectional Encoder Representations from Transformers (BERT) for contextualized topic modeling. Due to the limited length of tweets (maximum of 280 characters) and the frequent use of characters like  \# and @, traditional topic modeling techniques like latent data analysis (LDA) \cite{blei_dynamic_2006,hays_cultural_1998} tend to produce less intelligible results \cite{egger2022topic}. Instead, we use a topic modeling method that relies on clustering word embeddings. Word embeddings model the semantic context \cite{mikolov_efficient_2013} of words by mapping corpus terms in semantic space ``in which distance represents semantic association'' \cite{griffiths_topics_2007} (cited in \cite{angelov_top2vec_2020}). We used the BERTopic package \cite{grootendorst2022bertopic} to cluster BERT embeddings \cite{reimers_sentence-bert_2019} using Hierarchical Density-Based Spatial Clustering of Applications with Noise (HDBSCAN) \cite{campello_density-based_2013}. Similar documents in the corpus will be closer to each other. Each document will also be closer to words semantically closer to it. 

\paragraph{Topic Coherence} To find the optimum number of topics for the BERTopic model, we trained 25 different models by changing the minimum cluster size (an HDBSCAN hyperparameter that controls the size of the clusters) by increments of 10 at a time.\footnote{\url{https://hdbscan.readthedocs.io/en/latest/parameter_selection.html?highlight=min_cluster_size\#selecting-min-cluster-size}} The minimum cluster size of the first model was set at 15, and the last model was set at 255. The higher the minimum cluster size, the lower the number of clusters and identified topics. To identify the best model, we calculated the coherence score of each model using the Gensim \textit{CoherenceModel} feature \cite{rehurek_coherencemodel}. Coherence values have been found to be good in approximating human ratings of a topic model ``understandability'' \cite{roder_exploring_2015}.  We selected the model with the highest coherence score (0.39), with a minimum cluster size of 45, that produces a topic model with 71 topics.

\begin{figure}[h!]
    \centering
    \includegraphics[width=0.99\linewidth]{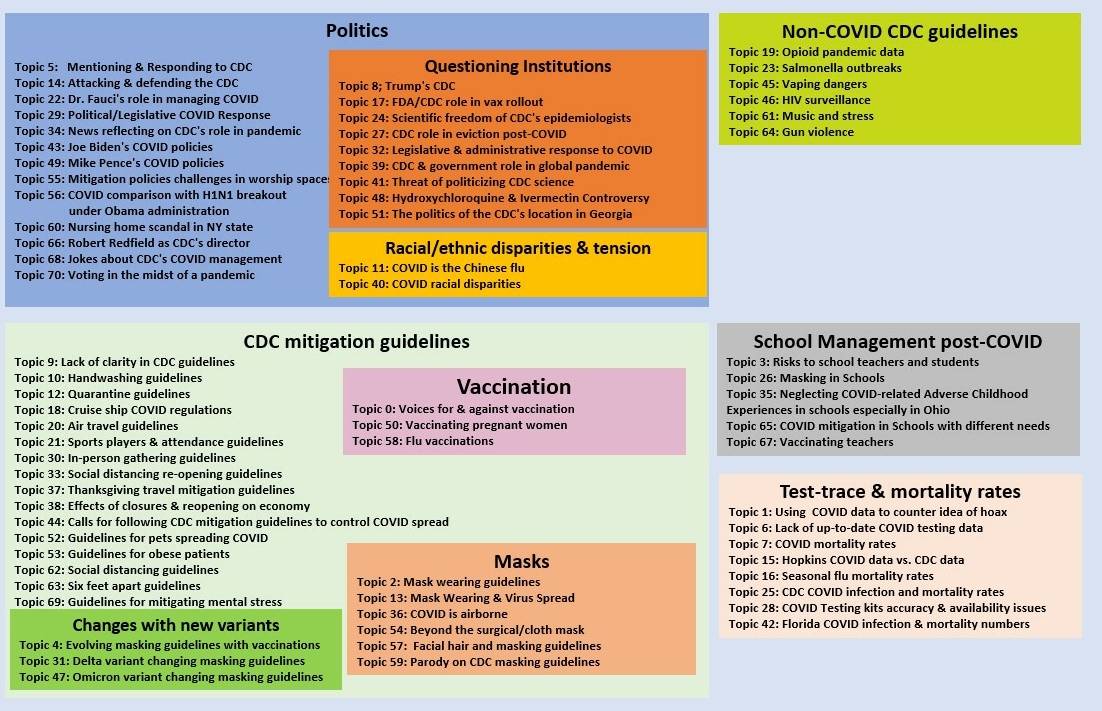}
    \caption{BERTopic clustering results grouped into five major themes, including COVID mitigation, partisan discourse, and public health equity. Themes reflect how discussions clustered semantically and ideologically.}
    \label{fig:top_themes}
\end{figure}

Figure \ref{fig:top_themes} shows the breakdown of topics into five main themes covering a similar discourse around COVID mitigation and politics. \S\ref{sec:meth_cluster}.  

\subsubsection{Sentiment analysis (N=1)} \label{sec:meth_sent}
To understand how the sentiments of tweets affected the way that people discussed topics in the CDC public, we used VADER, a sentiment analysis tool that is frequently applied to Twitter text \citep{elbagir2019twitter} which calculates the valence and polarity for each word (positive, negative, or neutral) along with the percentage of said polarity in a composite score. VADER labels text as having positive sentiment if the composite score is above 0.05, negative sentiment if the score is below -0.05, and neutral sentiment if the score is between -0.05 and 0.05 \citep{hutto2014vader}. 

\subsubsection{Credibility of Information Sources (N=1)} \label{sec:credibility}
We used the Iffy Index of Unreliable Sources,\footnote{\url{https://iffy.news/index/}} a public-facing dataset of low-credibility sources\footnote{\url{https://docs.google.com/spreadsheets/d/1ck1_FZC-97uDLIlvRJDTrGqBk0FuDe9yHkluROgpGS8/gviz/tq?tqx=out:csv&sheet=Iffy-news}} reviewed by Media Bias/Fact Check,\footnote{\url{https://mediabiasfactcheck.com/methodology/}} and was used to determine the credibility of medical data about COVID-19 vaccination information on Twitter \cite{pierri_2023}. This dataset contains the following 12 features:

\begin{enumerate}
    \item Domain: Site address without ``https:\/\/'' and ``www.'' \newline
    (e.g. google.com instead of https:\/\/www.google.com)
    \item Site rank: The Alexa rank of a site’s traffic
    \item Year online: The year the site was established
    \item Name: The site name
    \item Factual categorization: Very high, high, mostly factual, mixed, low, very low
    \item Bias (least biased/pro-science, right-center/left-center, left/right, questionable/conspiracy pseudoscience)
\end{enumerate}

The Iffy Index only includes sites with a low-credibility rating and categorizes as either Conspiracy/Pseudoscience (CP) or Questionable Source/Fake News (FN). 

\subsubsection{Twitter Platform Signals (N=9)} \label{subsec:platform_sig} Twitter provides affordances for propagating messages and expanding your audience. We will use the features identified in the section and listed below:
 \begin{enumerate}
        \item Retweet count (RT)
        \item Quote Tweet count (QT)
        \item Tweet likes (Like)
        \item Direct responses to Tweets (Rep)
        \item Rich Media: Tweet uses rich data (e.g., pictures or videos) 
        \item Verified account
    \end{enumerate}

\subsection{Exploring discourse in CDC Twitter publics}
To identify different publics within the CDC public, we will cluster users into different groups that share topics of discussion and user behavior.

\subsubsection{Clustering to find communities} \label{sec:meth_cluster}
Since our data is not labeled, we used the features presented earlier (topics, sentiment, the properties of the Twitter account) to cluster Twitter users in our dataset. Similar to methods adopted by Hussein et al. \cite{hussein2021cluster}, we used K-Means clustering on two feature types: 
\begin{enumerate}
    \item Discourse-specific features:
    \begin{enumerate}
        \item Average topic weights discussed by users (see \S\ref{sec:topics}) 
        \item Average user sentiment when discussing the topics identified in 1(a) (see \S\ref{sec:meth_sent})
        \item Credibility of sources used by Twitter users (see \S\ref{sec:credibility})
    \end{enumerate}
    \item User-specific features presented in \S\ref{subsec:platform_sig}
\end{enumerate}

\begin{figure}
    \centering
    \includegraphics[width=0.5\linewidth]{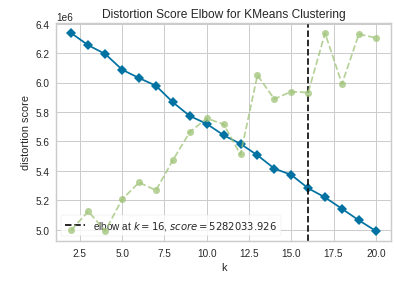}
    \caption{Using the elbow method, we identified 16 clusters as being the best number based on distortion error.}
    \label{fig:elbow_method}
\end{figure}

When using K-Means models, choosing the optimal K number of clusters is of central importance to the quality of the clusters. We used the Yellowbrick Python package \cite{bengfort_yellowbrick_2018} to select the optimal number of clusters using the knee point detection algorithm \cite{8405717}. The optimal k=16 is presented in Figure \ref{fig:elbow_method}.   

\subsubsection{Analyzing cluster content} To interpret discourse within each cluster, we combined computational and qualitative approaches rooted in discourse analysis. We used two methods to characterize the linguistic patterns of cluster users: Log Likelihood Ratio (LLR), which identifies distinguishing terms to define the “aboutness” of a discourse \cite{gupta_07}, and Tag2Vec, which uses document embeddings to find semantically similar words associated with cluster tags or sentiment framings \cite{chen2017doctag2vec}. These methods helped us surface salient terms in both tweet content and user biographies.

Our qualitative analysis followed a multi-step protocol. For large clusters, we sampled 100 users, including at least 10

Drawing from discourse theorists who argue that language does more than transmit information—it also shapes social realities \cite{vanDijk1985semantic, schiffrin1987discourse, potter1987discourse}—we treated tweets not merely as retweet or mention events but as meaningful communicative acts \cite{papacharissi2016affective, utz2013crisis}. In this view, discourse becomes a resource through which users structure knowledge and make sense of uncertainty \cite{vanDijk1985semantic}. Because topic models are not self-interpreting, we applied qualitative interpretive strategies to contextualize each of the 71 topics generated with the highest coherence scores \cite{chang_reading_2009, dou_paralleltopics_2011, park2020stakeholder, chancelor_et_al_18}.

This interpretive process involved three major steps:

    \begin{enumerate}
        \item Theme Identification: We randomly sampled 100 tweets from each topic to draft preliminary framings that reflected key thematic patterns.
        \item Sentiment Differentiation: Using LLR and Tag2Vec, we identified the top words associated with each theme by sentiment polarity (positive, negative, neutral). We sampled 100 tweets for each theme–sentiment pairing, generating a corpus of 6,000 tweets. Sentiment distributions were wide-tailed and highly polarized, consistent with prior work showing non-normal emotional distributions in crisis discourse \cite{kallner2017laboratory}.
        \item Consensus Coding: Team members collaboratively coded tweets and developed shared analytic memos. We discussed interpretations in weekly meetings to refine categories and identify overarching themes and tensions \cite{McDonald_et_al_19}.
    \end{enumerate}

In line with best practices in social media research ethics, we do not quote individual users verbatim unless they represent public institutions such as the CDC or major news outlets. This follows guidance from Fiesler and Proferes \cite{fiesler_participant_2018}, who argue that users do not expect academic reuse of their tweets, even if technically public. Instead, we use paraphrased examples and thematic summaries, consistent with Bruckman’s framework for protecting user anonymity online \cite{bruckman_studying_2002}. 

\subsubsection{Analyzing cluster structure}
We used Krackhardt and Stern’s \cite{krackhardt1988informal} “E-I” index \textemdash a measure of network insularity of cohesion. The index ranges between [-1,1]. The closer the value to +1, the more externally connected the network while the closer to -1, the higher the chance of an echo chamber being formed in the network \cite{wolfowicz2023examining,bruns2017echo}.  

The "E-I" index was built on stance (topic X sentiment) network built using mentions and responses as edges (reflecting the three propagation affordances in Twitter). 

\subsection{Temporal Analysis} \label{meth:time}
Earlier work also analyzed the temporal dynamics of online discourses and how they are affected by the onset of COVID \cite{lee2021using} by specifying one breaking point, March 11\textsuperscript{th}, which is when COVID was declared a pandemic. The authors then used the Chow test \cite{wooldridge2019introductory} to measure the significance of the change in topic-level discussions before and after that date. However, given that it is difficult to define the different phases of a health crisis beforehand (see \S\ref{related_work:flashpoint}), we describe the change-point detection method we used in \S\ref{meth:PELT}. 

\subsubsection{Temporal Change-point Detection} \label{meth:PELT}
We used the Pruned Exact Linear Time (PELT) algorithm for change-point detection to identify significant shifts in sentiment and topic prevalence over time within each discourse cluster. PELT is known for its conservative and accurate detection of temporal changes in time series data \cite{Truong2020,killick2012optimal,wambui2015power}, and has been applied in prior Twitter analyses, such as COVID-19 discourse \cite{valdez2020social}.

To interpret the direction and significance of trends between change-points, we applied the Mann-Kendall non-parametric trend test \cite{hussain2019pymannkendall}. In visualizations, red dashed lines indicate change-points; arrows show trend direction; and green dots mark statistically significant segments (p < 0.05).

\section{Findings} \label{sec:findings_1}
In this section, we present our key findings across five thematic areas. We begin by identifying patterns of echo chambers and cross-cutting discourse within the CDC Twitter publics. We then analyze how users responded to CDC mitigation guidelines, including shifts in messaging clarity and polarization. Beyond COVID, we examine discourse around other CDC guidelines, explore the politicization of public health communication, and track how marginalized communities—particularly those focused on racial equity—sustained engagement over time despite declining broader interest. Throughout this section, we analyze the mostly polarized discussions under each theme, and we introduce the different stakeholders and their respective \textcolor{blue}{positive}, \textcolor{purple}{neutral}, and \textcolor{red}{negative} communication frames. 

\begin{figure}[h!]
    \centering
    \includegraphics[width=1\linewidth]{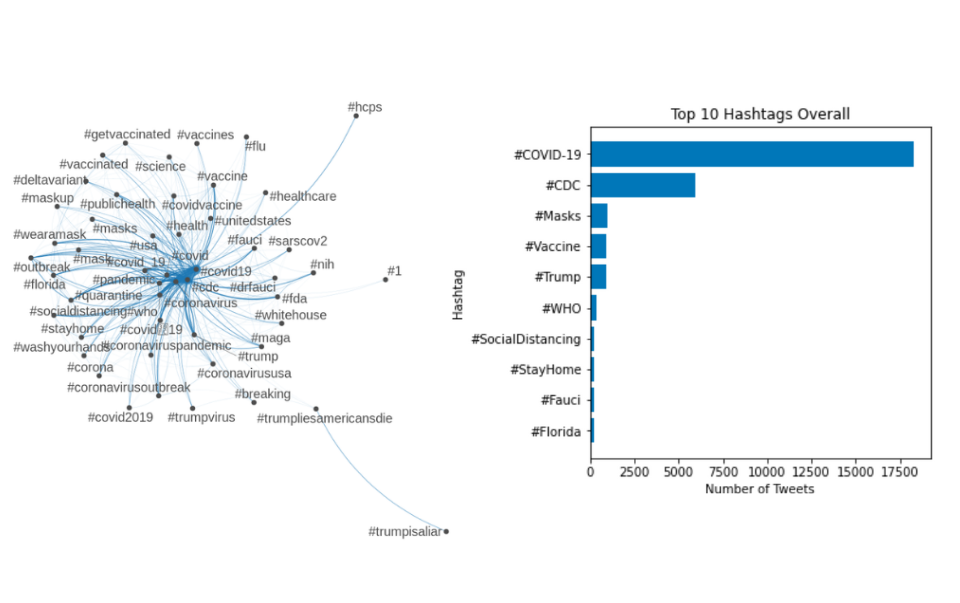}
    \caption{The figure on the left shows the co-occurrence of hashtags with the \#CDC. The figure on the right shows the top hashtags in the discourse. We collapsed similar hashtags together. For example, ‘\#coronavirus’, ‘\#covid19’, ‘\#covid-19’, ‘\#sars’, ‘\#coronaviruspandemic’, ‘\#pandemic’, etc. were collapsed to simply to ‘\#COVID-19’. ‘\#CDC’ and ‘\#cdc’ were collapsed to ‘\#CDC’. The hashtags ‘\#vaccine’, ‘\#GetVaccinated’, ‘\#vaccinated’, and ‘\#VaccinesWork’ were collapsed to ‘\#Vaccine’.  Likewise, ‘\#Trump’, ‘\#TrumpVirus’, ‘\#TrumpLiesAmericansDie’ were collapsed to ‘\#Trump’. Lastly, ‘\#Masks’, ‘\#WearAMask’, ‘\#MaskUp’, ‘\#masks’, and ‘\#mask’ were collapsed to ‘\#Masks’. }
    \label{fig:tophash}
\end{figure}

To get an overview of the discourse, we analyzed the top 10 hashtags and their co-occurrence network with \#CDC \cite{turker_2018} (Figure \ref{fig:tophash}). This revealed three main themes:

    \begin{enumerate}
        \item Mitigation – including hashtags related to masking (\#maskup, \#masks), vaccination (\#getvaccinated, \#covidvaccine), and lockdowns (\#quarantine, \#stayhome).
        \item Science and medical expertise – featuring \#NIH, \#WHO, \#FDA, \#DrFauci, and \#HCPCS (Healthcare Common Procedure Coding System), the latter linked to insurance-related discourse.
        \item Politics and public figures – centered on Trump and Fauci, with \#Florida prominent due to state-versus-CDC conflicts (\#MAGA, \#trumpvirus, etc.; see more detail in \S\ref{sec:covid_test}).
    \end{enumerate}

We now explore these discourse strands across different publics in more detail next.

\subsection{Many echo-chambers}
The E-I index analysis reveals distinct patterns of interaction across the 16 discourse clusters. Clusters 5 through 16 each have an E-I index of -1.0, indicating perfect echo chambers—users in these clusters interacted exclusively with others who shared the same stance, with no cross-stance engagement. Cluster 12 stands out with a moderately negative E-I index of -0.24, reflecting some interaction across stance lines but still largely internal communication. Similarly, Cluster 3 shows a slight echo chamber tendency with an E-I index of -0.19. In contrast, Clusters 2 and 1 exhibit the most cross-cutting discourse: Cluster 2 has a strongly positive E-I index of 0.75, while Cluster 1 is the most integrative, with an E-I index of 1.0, meaning all interactions occurred between users of different stances. Clusters 4, 10, and 14 could not be evaluated due to missing or insufficient interaction data. These results suggest that while many clusters are ideologically siloed, a few serve as bridging spaces where users engage across lines of disagreement.

\begin{table}[!ht]
    \small
    \centering
    \caption{Top topics and theme distributions for each discourse cluster, including major focus areas and retweet averages (part I)}
    \begin{tabular}{|l|p{2cm}|p{8cm}|l|}
    \hline
        \textbf{\#} & \textbf{Cluster Name} & \textbf{Topics and Weights (\%)} & \textbf{RT avg} \\ \hline
        \textbf{1} & COVID racial disparities & COVID racial disparities (73), Voices for and against vaccines (1), COVID testing kits accuracy \& availability (1), Using COVID data to counter idea of hoax (1), Hopkins COVID data vs. CDC data (1) & 7.78 \\ \hline
        \textbf{2} & Mitigation policy discussion & Using COVID data to counter idea of hoax (5), Mentioning and responding to CDC (5), Mask wearing guidelines (4), Jokes about CDC COVID management (3) & 10.85 \\ \hline
        \textbf{3} & Partisan discussion of COVID response & Legislative \& administrative response to COVID (25), Trump's CDC (8), Threat of politicizing CDC science (5), Mentioning \& responding to CDC (5), Attacking and defending the CDC (5) & 1.41 \\ \hline
        \textbf{4} & Guideline changes with Omicron Variant & Omicron variant changing masking guidelines (78), Voices for \& against vaccines (3), Hopkins COVID data vs. CDC data (1), Six feet apart guidelines (1), Air travel guidelines (1) & 10.49 \\ \hline
        \textbf{5} & CDC & Voices for \& against vaccination (11), Using COVID data to counter idea of hoax (4), Flu vaccinations (3), Salmonella outbreaks (2), Risks to school teachers and students (2) & 26K \\ \hline
        \textbf{6} & Pro-mitigation & Parody of CDC masking guidelines (8), Voices for and against vaccinations (2), Mask wearing guidelines (2), Calls for following CDC mitigation guidelines to control COVID spread (1) & 1.53 \\ \hline
        \textbf{7} & No clarity in CDC guidelines & Lack of clarity in CDC guidelines (49), Calls for following CDC mitigation guidelines to control COVID spread (17), Mentioning and responding to CDC (1), Voices for \& against vaccination (1), Attacking \& defending CDC (1) & 0.29 \\ \hline
        \textbf{8} & Obesity and COVID & Guidelines for obese patients (79), Voices for \& against vaccination (2), Delta variant changing masking guidelines (1), Mentioning \& responding to CDC (1), Hopkins COVID data vs. CDC data (1) & 0.84 \\ \hline
        \textbf{9} & Gun violence & Gun violence (80), Voices for \& against vaccination (2), Jokes about CDC's COVID management (1), COVID racial disparities (1), Delta variant changing masking guidelines (1) & 6.3 \\ \hline
        \textbf{10} & Food-borne diseases outbreak & Salmonella outbreak (71), Mentioning \& responding to CDC (1), Lack of clarity in CDC guidelines (1), Voices for and against vaccination (1), Mask wearing guidelines (1) & 2.43 \\ \hline
        \textbf{11} & Vaccinating pregnant women & Vaccinating pregnant women (72), Using COVID data to counter idea of hoax (2), Mentioning \& responding to CDC (2), Voices for \& against vaccination (1), Attacking and defending the CDC (1) & 9.03 \\ \hline
        \textbf{12} & School sports \& community events & Voices for \& against vaccination (42), Sports players \& attendance guidelines (4), In-person gathering guidelines (3), Social distancing re-opening guidelines (3), Florida COVID infection \& mortality numbers (2) & 4.91 \\ \hline
    \end{tabular}
    \label{tab:Cluster_topics}
\end{table}

\begin{table}[!ht]
    \small
    \centering
    \caption{Top topics and theme distributions for each discourse cluster, including major focus areas and retweet averages (part II)}
    \begin{tabular}{|l|p{2cm}|p{8cm}|l|}
    \hline
        \textbf{\#} & \textbf{Cluster Name} & \textbf{Topics and Weights (\%)} & \textbf{RT avg} \\ \hline
        \textbf{13} & Air travel guideline changes & Air travel guidelines (77), Using COVID data to counter idea of hoax (1), Social distancing re-opening guidelines (1), CDC COVID infection and mortality rates (1) & 1.24 \\ \hline
        \textbf{14} & Mask wearing guidelines & Facial hair and masking guidelines (80), Voices for \& against vaccination (2), COVID testing kits accuracy \& availability issues (1), Evolving masking guidelines with vaccinations (1), Calls for following CDC mitigation guidelines to control CDC spread (1) & 0.76 \\ \hline
        \textbf{15} & School reopening dynamics & Risks to school teachers \& students (75), Voices for \& against vaccination (1), Vaccinating teachers (1), Neglecting COVID-related adverse childhood experiences (1), Mentioning and responding to CDC (1) & 4.4 \\ \hline
        \textbf{16} & Quarantine dynamics & Quarantine guidelines (77), Using COVID data to counter the idea of a hoax (1) Voices for \& against vaccination (1), Lack of up-to-date COVID testing data (1), Dr. Fauci's role in managing COVID (1) & 1.22 \\ \hline
    \end{tabular}
    \label{tab:Cluster_topics}
\end{table}

\subsection{CDC mitigation guidelines} \label{sec:cdc_mitigation}
In this theme, discourse focused on different mitigation policies the CDC has called for. These include the following: (1) masks; (2) vaccines; (3) social distancing, and (4) COVID test-trace. Discussions about these topics were polarized between those who thought CDC guidelines were not stringent enough and those who did not believe the CDC had the right to mandate any guidelines. They revolved around discussing and clarifying CDC mitigation guidelines with the introduction of new variables, such as variants of interest (e.g., the Delta and Omicron variants), as well as the evolving understanding of the efficacy of vaccines.

\subsubsection{Mitigation guidelines in the social context}
Discourse about the validity of CDC guidelines and their effectiveness was a major area of discussion. Cluster 2 (see Table \ref{tab:Cluster_topics}), one of the bigger clusters (with about 40,000 Twitter users), which contained almost 2,000 verified users, focused on discussing how COVID is not a hoax. Many of those leading this charge were verified public health professionals (N=23) who explained why CDC mitigation guidelines (e.g., masking and social distancing) are important to control the spread of the disease. Others (N=17) were identified as reporters, anchors, and medical journalists who reported on the initially evolving mitigation guidelines presented by the CDC. However, some users who self-identified as conservative and/or Trump supporters (N=5) questioned the veracity of the pandemic, and many considered it a hoax and used concepts like ``plandemic'' to explain it away. Those views were sometimes in response to verified public health accounts. Others mocked anyone who ever masked outdoors, even at the height of COVID surges. The frustrations expressed by anti-masking users were couched in the idea of freedom - a word that often appeared in Twitter user descriptions in Cluster 2. Another frame pushed by most users in cluster 8, most of them self-identifying as politically conservative (N=4), question the negative health outcomes from COVID itself. They instead argue that underlying conditions, specifically obesity, are to blame for high death rates. In other words, obesity killed people who happened to have COVID.

Cluster 14 focuses on guidelines for where masks need to be worn (e.g., spaces like grocery stores, hospitals, etc.) and how to wear masks correctly, especially referencing facial hair for men and makeup for women (cosmetology is one of the identifying words in Table \ref{tab:Cluster_users}). The efficacy of the surgical masks is brought into question. This discourse is about N95/KN95 masks and how they are more efficacious than other masks, especially with newer variants. Several public health professionals criticized the CDC guidelines about cloth and N95 masks. Specifically, they were arguing for mandating N95 masks in hospitals and for vulnerable populations. They were also frustrated with the scarcity of N95 masks and other protective gear in hospital settings. To make their point, the opponents of masking mandates posted earlier CDC guidelines that did not recommend masks along with the current recommendations that say otherwise. This is one of the earlier examples of the lack of clarity in CDC guidelines. Three of the users relied on screenshots of earlier CDC guidelines to argue against new, usually stricter, mitigation efforts (e.g., you should wear a mask even if you are not symptomatic).  

In Cluster 12, these concerns were localized to school district sports activities and how they can be reopened safely for families, teachers, and student-athletes. While some parents were pushing for a return to "normal," medical professionals pointed out that COVID infections of youth athletes can have long-term effects on them. They cautioned against reopening sporting events without adequate mitigation, including COVID testing. Many accounts included high school and college athletic departments (N=11), coaches and heads of athletic teams (N=5), and school administrators (N=2).

\paragraph{\textcolor{blue}{Positive Communication Frame: Wearing a mask is important to protect yourself and others}}
\paragraph{\textcolor{purple}{Neutral Communication Frame: KN-95 masks give a higher level of protection than cloth masks}}
\paragraph{\textcolor{red}{Negative Communication Frame: Forcing people to wear a mask goes against our freedoms}}

\subsubsection{As the pandemic evolves, guideline clarity devolves}
Cluster 7 is focused on the idea that CDC guidelines are not clear. The discourse in this cluster was mostly pro-mitigation and leaned on the liberal side politically. For example, in Table \ref{tab:Cluster_users}, we can see that words like \#ANTIFA, \#BlackLivesMatter, etc. were indicative of the biographies in this cluster. The focus on professional status was also important in this cluster (see words like scholar, chief medical resident etc.). Medical professionals (N=17) wanted the CDC to issue simpler guidelines and not change them in response to subtle changes in the number of COVID infections. Many also pointed out that the CDC should use other metrics besides death rates because the medical and scientific community still has little understanding of long-term effects of COVID infections. The change in quarantine guidelines was critiqued by vascular and pediatric medical doctors in Cluster 16. Specifically, they were concerned about the shortening of the quarantine period from 14 days to five days. This issue was also picked up by different stakeholders in Cluster 13. While reporters were presenting the CDC guideline changes, epidemiologists (N=4) and other healthcare professionals (N=8) as well as medical institutions (N=5) argued against this change. They believed five days did not provide enough protection against spreading COVID to others. However, this was countered by other stakeholders, including the Air Line Pilots Association and travelers, who professed to have lost patience with all the new travel guidelines. In a related way, many posters were excited about CDC guidelines that do not require tests or masking on airplanes or indoors if fully vaccinated. Some medical professionals (N=5) and advocates (N=4) still encouraged masking to protect themselves and others. There was disappointment in the new guidelines and worry among essential workers about un-vaccinated individuals potentially exposing them to the virus. As masking guidelines were changing, questions arose about the efficacy of the vaccine in stopping infections. These questions emerged especially as the vaccine became less effective after the Delta and Omicron variants (Cluster 4). Some of the discourse touched on reintroducing masking, especially at peak COVID infection points (e.g., the Omicron wave in late 2021). Another point of friction focused on the changes that came with re-opening schools. 

\paragraph{\textcolor{blue}{Positive Communication Frame: While the vaccine will not protect you from the Omicron variant, it still protects from severe illness}}
\paragraph{\textcolor{purple}{Neutral Communication Frame: Because the new variants are different from the wild-type COVID, vaccines are less effective against them}}
\paragraph{\textcolor{red}{Negative Communication Frame: Why did the CDC say that the vaccine was over 90\% effective if it does not protect against the new variants?}}

\subsubsection{School management post-COVID}
Cluster 15 is focused on school openings. The discussions in this cluster are split between those pushing for opening schools back up (N=10), three of whom self-identified as parent advocates. One of the parent advocates argue school closures have negatively affected the managing of adverse childhood experiences (ACE) since many of the mandatory reporters (e.g., social workers, teachers, school councilors, etc.) do not have contact with children. Teachers (N=7), superintendents (N=3), and teacher union leaders (N=1) called for teachers to have a higher priority on the vaccination list to potentially allow for faster school re-openings, which made many parents happy. However, since the vaccines were not approved for children at the time, those same accounts were worried about COVID being transmitted between students. Therefore, these accounts called for masking in schools regardless of vaccination status. At the same time, many of the advocate parents that were introduced earlier also perceived masking as unsafe and unfair to children. 

\paragraph{\textcolor{blue}{Positive Communication Frame: Masking, which protects both teachers and students, is more important as new variants appear}}
\paragraph{\textcolor{purple}{Neutral Communication Frame: These are the precautions that players and parents need to take for high school sports events to continue}}
\paragraph{\textcolor{red}{Negative Communication Frame: Closing schools and forcing students to mask are both affecting students negatively}}

\subsection{Non-COVID CDC guidelines} \label{sec:cdc_non_covid}
While most of the discourse in our dataset has focused on COVID, there were other issues that the CDC continued to manage at the time. For example, veterans and pain advocates discussed the need for relaxing CDC rules for opioids, especially for veterans who might have particular needs based on their injuries. They also noted that isolation from the COVID shutdown has added to the pain and mental health challenges faced by veterans. Some of the accounts identified as pain advocates (N=7), many of whom were veterans (N=11) connecting with others who share similar challenges, highlight how their pain was especially exacerbated by the COVID lockdown and pressures that affected their peer support group activities.

As Table \ref{tab:Cluster_users} shows, words like MAGA, \#Liberal, \#GunReformNow, \#GunlawsSaveLives, and gun-related trauma are all representative of the way users in Cluster 9, which focuses on the polarizing issue of gun control, described themselves. While gun reform was the focus, many self-identified libertarians (N=3) framed their freedom and belief in the Second Amendment as their rejection of CDC mandates. On the other hand, advocate organizations like Moms Demand Action, which are interested in public safety through gun law reform, framed CDC guidelines as central to community safety. One of the less polarized clusters was Cluster 10, which revolved around food-borne diseases (e.g., Salmonella) about which the CDC continued to report. 

\paragraph{\textcolor{blue}{Positive Communication Frame: Remembering my sister who passed one year ago today that happens to be \#opioidawarenessday. You can find Narcan and social support at our center if you need help [address]}}
\paragraph{\textcolor{purple}{Neutral Communication Frame: The CDC is providing data showing that fentanyl is the \#1 cause of death in Americans age 18-45. Plans need to be made to address this issue.}}
\paragraph{\textcolor{red}{Negative Communication Frame: The \#OpioidHystera that has been led by the CDC and has caused severe pain patients to be treated like criminals.}}

\begin{table}[!ht]
    \small
    \centering
    \caption{Cluster metadata including user counts, verified status, top biography descriptors (doc2vec and LLR), and E-I index. These indicators reflect ideological cohesion and engagement diversity within each cluster (part I).}
    \begin{tabular}{|l|p{2cm}|p{1cm}|p{3.5cm}|p{3.5cm}|p{1.5cm}|}
    \hline
        \textbf{\#} & \textbf{Cluster Name} & \textbf{Users} & \textbf{Top desc doc2vec} & \textbf{Top desc LLR} & \textbf{E-I Index} \\ \hline
        \textbf{1} & COVID racial disparities & 121 (13) & @georgetown, displaced, self-proclaimed, care, oncology, \#healthequity, alliance, wellness, Latinx, \#Harmreduction, chicano, \#Goodtrouble, ADOS & law, latino, black, program, medicine, POTUS, community, gender, socially, racist, violence, getvaccinated, epidemiologist, care, feminist, adjunct, diversity, researcher & 1.00 \\ \hline
        \textbf{2} & Mitigation policy discussion & 39,749 (1,932) & first amendment,\#ANTIFA, \#BuildTheWall, \#ClimateAction, \#CONSERVATIVE, \#RightMatters, \#AllLivesMatter, \#Metoo, \#MAGA, Supporter, \#Truthmatters & freedoms, dweller, proponent, bioethics, thedemocrats, Johns Hopkins University, Come get them  & 0.75 \\ \hline
        \textbf{3} & Partisan discussion of COVID response & 4,918 (93) & \#patriot, \#bidenharris2020,\#independent, GoBucks, \#StrongerTogether, \#Science, Realist, \#TeamPelosi, Latter-Day, \#BlueWave, \#VoteBlueNoMatterWho & retired, trump, resist, donald, steelernation, detest, theresistance, equalrights, hamiltonmusical, election, angry & -0.19 \\ \hline
        \textbf{4} & Guideline changes with Omicron Variant & 41(6) & Educator, @Harvard, Educated, \#books, Forbes, @ClemsonUni, lecturer, mgmt, Epidemiology, Strategist, CEO, @HowardU & njspotlightnews, action10news, cichealth, byline, columbiamsph, kennedy\_school, columbiamed, newscaster, nyuniversity & --- \\ \hline
        \textbf{5} & CDC & 1(1) & NA & privacy, centers, cdc, updates, source, control, prevention, daily, disease, safety, official, health & -1.00 \\ \hline
        \textbf{6} & Pro-mitigation & 91(1) & \#Geek, OPINIONS, Makeup, Thinker, \#DubNation, \#RedSox, \#Braves & long distance relationship, nonconformist, hospitalityindustry, card & -1.00 \\ \hline
        \textbf{7} & No clarity in CDC guidelines & 1,280(40) & scholar, \#ANTIFA, hacker, \#SaveOurDemocracy, \#BlackTransLivesMatter & fanboys, relax, wtvc, dogfather, chiefmedicalresident, finanacialprofessional & -1.00 \\ \hline
\hline
    \end{tabular}
    \label{tab:Cluster_users}
\end{table}

\subsection{Politics} \label{sec:politics}
School management was not the only area of the pandemic response that was politicized. In this section, we will show how Twitter users focused on the lack of governmental preparation that was evident in the lack of COVID tests. We follow this with a wider discussion of government response (or lack thereof) to the pandemic. Finally, we focus on racial disparities in the healthcare system in general and through the COVID prism in particular.

\subsubsection{Test-trace and mortality rates} \label{sec:covid_test}
The lack of COVID test kits, especially in the early days of the pandemic, is central to this theme. As seen in Table \ref{tab:Cluster_users}, the accounts of public health professionals (N=5), medical research institutions (N=2), and self-identified liberal politicians (N=3) engaged in critiquing, and at times directly attacking, the Trump administration's management of testing. The main issue they focused on was that accurate reflection of infection counts is important for safeguarding others, especially in healthcare settings (e.g., \#nurselife). They used hashtags like \#wherearethetests, \#liesaredeadly, and \#incompetentinchief. The focus on the lack of testing and questioning mortality rates was associated with negative Tweets as discussed in \S\ref{subsec:negative}. There was a specific focus on the COVID testing strategy in Florida where the @HealthyFla (Florida Dept. of Health) determined that only symptomatic people would be tested for COVID as opposed to testing a larger sample of the population to assess whether lock-downs are needed. Governor Desantis' account (@govrondesantis) is frequently mentioned in this discussion as many point to his influence in this COVID testing strategy. Given how politicized the testing theme has been, we will explore politicization of the discourse further next.

\paragraph{\textcolor{blue}{Positive Communication Frame: While there is currently a lack of COVID testing kits, supply chains are being created}}
\paragraph{\textcolor{purple}{Neutral Communication Frame: It is currently difficult to measure the number of COVID deaths because of the lack of tests}}
\paragraph{\textcolor{red}{Negative Communication Frame: Not having an adequate supply of tests is a direct consequence of bad planning by the CDC and the administration}}

\subsubsection{Politicization of COVID response} \label{subsec:results_publics_politicization}
In Cluster 3, the discourse focused on the threat of the politicization of COVID. At the outset of the pandemic, the CDC was frequently labeled as ``Trump's CDC'' because the pandemic budget was slashed just in time for the COVID pandemic, which deeply affected the preparedness of the American public health system. Many of the users in this cluster used signals like \#VoteBluenoMatterWho, \#Resist, \#TheResistance, \#BlueWave, and \#BidenHarris. Three users self-identified as housing advocates of a moratorium on evictions, which was eventually invoked by the CDC.

There is a more positive view of the vaccination roll-out plan as more people receive the vaccine and the focus moves from lock-downs and pressure on the healthcare system. However, public health researchers (N=4) and reporters (N=6) addressed how COVID had a ``racial bias,'' as its effects on racial minorities were significantly worse without any response to this health outcome bias from state institutions. We focus on racial disparity in more detail next.

\paragraph{\textcolor{blue}{Positive Communication Frame: Thank you to the CDC, to the scientists, and to government officials who created the policies that allowed me to get the vaccine so I can go back to my normal life!}}
\paragraph{\textcolor{purple}{Neutral Communication Frame: CDC advising insurance companies that PrEP and similar medications should be fully covered under current insurance plans! This is imperative for HIV prevention.}}
\paragraph{\textcolor{red}{Negative Communication Frame: People are dying and COVID is spreading even faster! Why is Food and Drug Administration (FDA) pushing its meeting to certify the vaccine? }}

\subsubsection{Racial disparities in a pandemic} \label{subsec:racial_disp}
Cluster 1 focused on health equity for Black, Latino, and Chicano populations by self-proclaimed advocates from these communities (N=7). These include non-profit organizations (N=3), such as the American Descendants of Slavery (ADOS) and the National Association for the Advancement of Colored People (NAACP), academic institutions (N=11), and faculty (N=9) who researched different areas of racial disparities, such as lack of access to vaccines or their higher exposure to the disease as front-line workers.  Similarly, medical institutions and scholars advocated for more equitable access to vaccines for pregnant women of racial minorities using hashtags like \#HealthEquity and \#VaccinesSaveLives. This framing focused on how minorities, especially because they serve in front-line jobs that were deemed essential, bore the brunt of COVID infections, and they suffered higher death rates and other negative health outcomes. The discussions called for universal health coverage and better access to resources (e.g., PCR testing), especially for members of minority groups. While discourse focusing on the intersection of race and healthcare decreased overall in the CDC public, cluster 1 continued its focus on Topic 40 (COVID racial disparities) throughout the period of the study (see \S\ref{sec:evolve_over_time}).

\paragraph{\textcolor{blue}{The NAACP is thankful for the equitable roll-out throughout the state. Learn where you can get your vaccine [through this link] to protect yourself and your family.}}
\paragraph{\textcolor{purple}{Neutral Communication Frame: The CDC says racism is a threat to public health outcomes. This is a good step to addressing health inequities.}}
\paragraph{\textcolor{red}{Negative Communication Frame: While local health departments are tracking COVID infections and deaths without grouping by race we cannot alleviate structural racism in public health.}}

\begin{table}[!ht]
    \small
    \centering
    \caption{Cluster metadata including user counts, verified status, top biography descriptors (doc2vec and LLR), and E-I index. These indicators reflect ideological cohesion and engagement diversity within each cluster (part II).}
    \begin{tabular}{|l|p{2cm}|p{1cm}|p{3.5cm}|p{3.5cm}|p{1.5cm}|}
    \hline
        \textbf{\#} & \textbf{Cluster Name} & \textbf{Users} & \textbf{Top desc doc2vec} & \textbf{Top desc LLR} & \textbf{E-I Index} \\ \hline
        \textbf{8} & Obesity and COVID & 87(2) & Kamala, Streamer, hcaker, \#research, conservative, stupid & obesity, ASMBS, nutrition, dietitian, lbs, biostatician, diabetes, investigating, sugar & -1.00 \\ \hline
        \textbf{9} & Gun violence & 76(4) & MAGA, vaccinated, Montanan, \#Riseup, \#GunReformNow, \#Liberal & securitywarriors, gunlawssavelives, endoflifeoptions, stopliars, votingrightsforall, trauma, momsdemand, 2nd & -1.00 \\ \hline
        \textbf{10} & Food-borne diseases outbreak & 278(11) & Meat, Pastry, Trash & bidenapocalypse, material, consumers & --- \\ \hline
        \textbf{11} & Vaccinating pregnant women & 70(12) & cweight, news10, childbirth, alabamastateu, newberrylibrary, iumedschool, mayocliniccv, umnmedschool, mildly & @sloan\_kettering, @EmoryUniversity, @News\_8, \#HealthEquity, \#VaccinesSaveLives, Neuroscientist & -1.00 \\ \hline
        \textbf{12} & School sports \& community events & 7760(584) & accredited, supporter & reporter, anchor, story, fox4kc, hiv, endorsements, uclaathletics, school, community, pediatric, comorbidity & -0.24 \\ \hline
        \textbf{13} & Air travel guideline changes & 209(23) & Air Line Pilots Association, vacation, cbsaustin, thedailybeast & @News\_8, @NY1, Nuggets, Toronto, Correspondent, @sloan\_kettering, @UCF, @HarvardMed, Epidemiology & -1.00 \\ \hline
        \textbf{14} & Mask wearing guidelines & 80(6) & FAU, Nuggets, @HofstraU & cosmetology, \#itsthesmallthings, bloggers, linesman, medtechs, imresidency, nyuwagner & --- \\ \hline
        \textbf{15} & School reopening dynamics & 997(53) & daughter, fall, accredited, childhood, counselor, curriculum, practitioner, athletic, vascular & teacher, schools, educator, learner, education, superintendent, principal, mom, district, business owner, association & -1.00 \\ \hline
        \textbf{16} & Quarantine dynamics & 134(8) & feet, tired, meaning & rights, vol\_baseball, vamosgalaxy, annmediasports, childhoodcancerawareness, oxy\_football & -1.00 \\ \hline
    \end{tabular}
    \label{tab:Cluster_users_cont}
\end{table}

\subsection{Change Over Time: Comparing Marginalized and Broader Publics} \label{sec:evolve_over_time}
The two PELT-detected time series reveal distinct temporal patterns in race-related discourse across the Twitter public (top) and Cluster 1 (bottom), which centers on COVID-related racial disparities. While race discourse in the broader public shows a sharp early decline, Cluster 1 maintains focused engagement with racial issues throughout the pandemic, as detailed in \S\ref{subsec:racial_disp}. This shows that flashpoints detected on the community level could be different than those that represent a specific set of stakeholders forming an interest public of their own - in this case, a healthcare equity public. 

In the top panel, the time series shows a prominent early concentration of race-themed content, with a statistically significant changepoint detected on February 25, 2020, marking a sharp and sustained decline in discourse. The period before this changepoint (highlighted in blue) is characterized by high variability and frequent spikes, while the period after (highlighted in pink) reflects a flattened trend with minimal engagement—suggesting that race was a prominent topic early on but quickly diminished in focus within this cluster as the pandemic took hold. In contrast, the bottom panel reveals a more dynamic and episodic pattern, with multiple changepoints spanning from April 2019 to April 2021. This cluster displays intermittent but sharp increases in race-themed engagement—particularly around mid-2020—corresponding to periods of heightened racial awareness and public events, such as the George Floyd protests and subsequent public health debates. The alternating pink and blue segments indicate bursts of discourse rather than a consistent trend, suggesting that users in this cluster engaged with race themes in reaction to external events rather than as a sustained topic. We will focus on the discourse to provide examples for discourse in both. 

"The first changepoint occurred between August 20, 2020 and October 24, 2020. The tweets in this breakpoint cluster around four intersecting themes that highlight how COVID-19 has deepened long-standing racial inequities in U.S. public health. \textbf{Racial Disparities in COVID-19 Mortality and Hospitalization} emphasize CDC data showing that Black, Hispanic, and Latinx individuals—especially children—face disproportionately high rates of COVID-19-related death and hospitalization. These tweets stress that such outcomes are not incidental but patterned and predictable. \textbf{Intersection of Race, Age, and COVID-19 Impact} highlights how younger adults, particularly those aged 25–44 from Hispanic and Latino communities, are also suffering disproportionately from the pandemic, challenging the perception that only older adults are vulnerable. \textbf{Critique of Systemic Racism in Public Health} brings attention to long-standing inequalities—such as drastically shorter life expectancies for Black individuals with Down syndrome compared to their white counterparts—and points to the broader discrediting of institutions as they become more racially diverse, exposing implicit racial bias. Finally, \textbf{Calls for Political and Structural Accountability} use CDC statistics as a foundation for civic and legislative demands. These tweets urge political leaders and voters to acknowledge and act on racial disparities, pushing for reforms that confront systemic injustice through policy change, taxation, and institutional responsibility.

The second changepoint was detected between April 2, 2021 and April 27, 2021. The discourse at this breakpoint reveals four interconnected themes centered on public health equity, racial justice, and institutional accountability. \textbf{Data Justice \& Visibility} captures how several users challenge public health institutions—particularly the CDC—for aggregating health data in ways that obscure disparities. Tweets emphasize that grouping Native Hawaiian and Pacific Islander (NHPI) populations with all Asians, or generalizing Indigenous health statistics, erases the distinct health challenges these communities face. \textbf{Public Health Advocacy for BIPOC Communities} reflects coordinated efforts to promote COVID-19 vaccination within communities of color—especially among Latinos and Hispanics—using CDC data to highlight disproportionate infection rates. These messages are often accompanied by hashtags like \#Vaccinate4All and endorsements from culturally aligned organizations such as @SaludAmerica and @NHMAmd. \textbf{Institutional Racism \& Accountability} emerges as users share personal experiences and collective critiques of medical racism, calling on public health entities like the CDC and OSU Wexner Medical Center to acknowledge and address ongoing harm—particularly to Black women—and to end institutional gaslighting. Finally, \textbf{Symbolic Actions vs. Structural Reform} critiques the disjuncture between public-facing declarations—such as the CDC’s recognition of racism as a public health threat—and the lack of deep structural change across health and policy institutions. While symbolic gestures are noted, users call for more meaningful accountability and systemic reform.

\begin{figure}[ht]
    \centering
    \includegraphics[width=1.0\linewidth]{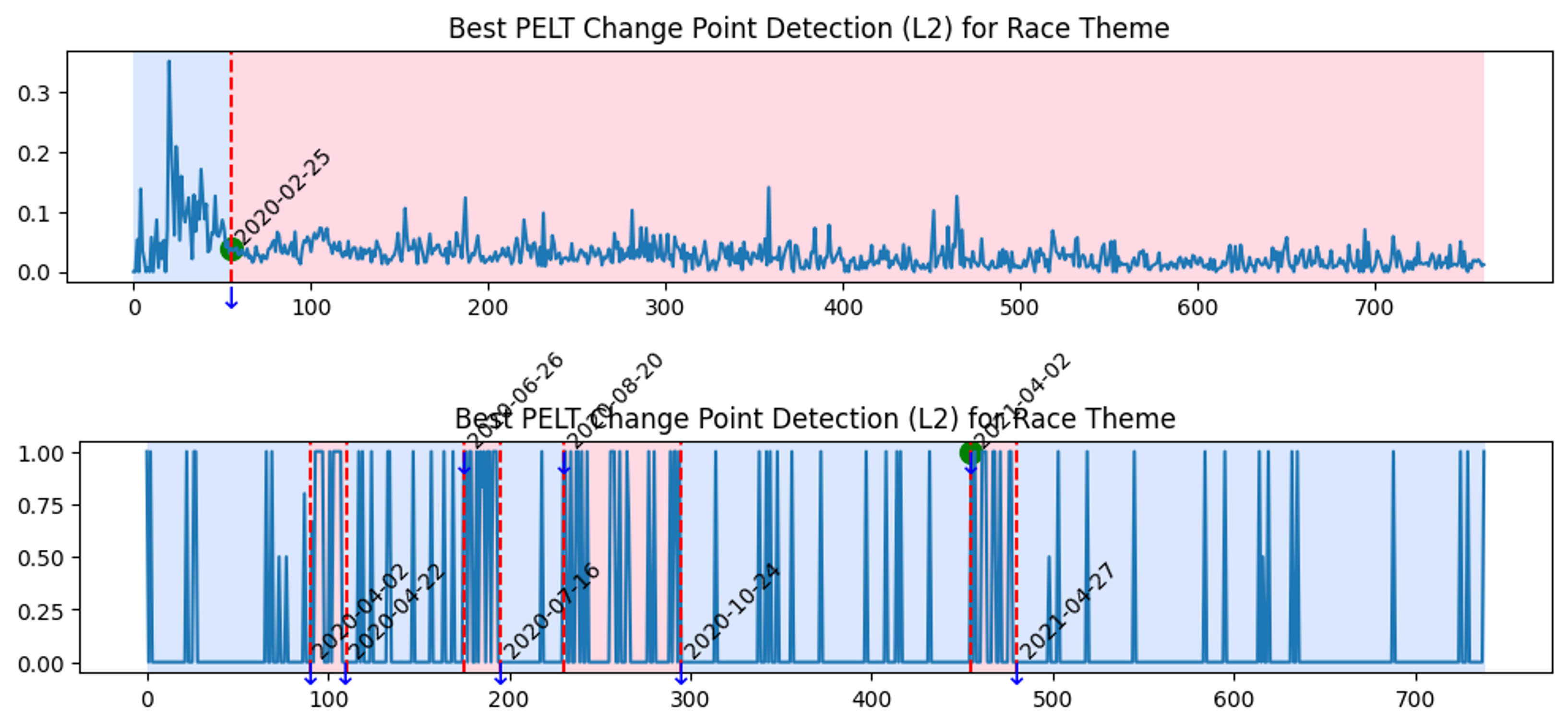}
    \caption{Time-series of race-related discourse across the full Twitter public (top) and Cluster 1 (bottom). Early decline in race discourse contrasts with episodic but sustained engagement by Cluster 1. Detected change-points (red lines) show how this marginalized public responded dynamically to external events, unlike broader publics that deprioritized racial equity over time.}
    \label{fig:race_cluster}
\end{figure}

\section{Discussion}
In this section, we reflect on the implications of our findings for understanding how crisis discourse unfolds across fragmented digital publics. In \S\ref{sec:earlier_work}, we reflect on how our findings relate to earlier work. In \S\ref{sec:engage_gen_ai}, we present design recommendations to better support both crisis organizations like the CDC and other publics. Finally, we provide policy recommendations for crisis organizations as they create and maintain long-term engagements with different publics (see \S\ref{sec:policy_rec}).  

\subsection{Reflecting on Earlier Literature} \label{sec:earlier_work}
To interpret the communication landscape that emerged around the CDC during the COVID-19 pandemic, we begin by unpacking the affective and ideological dynamics that shaped public engagement. In Section \S\ref{subsec:context}, we examine how partisan identity and political framing contributed to the formation of polarized discourse clusters that often operated in ideological isolation. In \S\ref{subsec:stakeholder}, we shift focus to reflecting on the spectrum of stakeholders and emotionality, exploring how affect—particularly anger, fear, and sarcasm—served both as a mobilizing force and a barrier to sustained cross-cluster dialogue. Finally, in \S\ref{subsec:marginalized_pub}, we analyze the structural configuration of discourse networks, highlighting how most clusters functioned as echo chambers while only a few enabled cross-cutting communication. Together, these subsections show how pandemic-related discourse was shaped not only by the content of messages but by the affective, ideological, and structural conditions under which they circulated.
\subsubsection{Crisis Communication in Context: Reflecting on Frames and Affective Polarization} \label{subsec:context}
Our findings reinforce prior research on the centrality of framing in public health crisis communication, especially when mediated through social media platforms. The CDC public on Twitter emerged as a highly polarized, affect-laden environment where hashtags served as tools of alignment and boundary marking. As with earlier work on issue and affective publics \cite{papacharissi2016affective, bruggemann2023debates}, users participated in identity-driven framing of core topics such as masking, vaccination, and school reopening. These frames often manifested in polarized pairings: public health versus personal liberty, science versus politics, and equity versus neutrality. This duality was most starkly evident in how users interpreted CDC messaging---either as lifesaving guidance or authoritarian overreach.

Emotion, or affect, played a critical role in shaping these interpretations. Hashtags functioned as affective signposts, clustering users into emotionally resonant communities. For example, frames like \#VaccinesSaveLives invoked collective responsibility, while \#VaccineRisk or \#Plandemic expressed fear and distrust. As found in prior work \cite{jin2010making, kim2011emotions}, messages that connected with pre-existing values and affective identities gained the most traction. Thus, affect did not just shape response but served as a filter through which users interpreted public health information.

\subsubsection{Stakeholders Across Publics: A Spectrum of Engagement} \label{subsec:stakeholder}
Our analysis of the CDC Twitter public revealed diverse stakeholder groups, which aligned closely with Hallahan's typology \cite{hallahan2000inactive}. Active stakeholders---such as healthcare professionals, teachers, and advocacy organizations---were visible in clusters focused on mitigation strategies and racial health equity (e.g., Clusters 1 and 2). Aware and aroused stakeholders, such as parents concerned about school reopening or travel guideline changes, populated Clusters 12 and 13. Inactive or minimally engaged users were harder to trace but may exist outside the dominant echo chambers.

Significantly, our findings demonstrate that different publics emerged not only around ideological lines but also around topical and affective interests. Cluster 1, for example, focused on racial health disparities and exhibited the highest level of cross-cutting engagement (E-I Index of 1.0), unlike most other clusters that were ideologically siloed. This cluster illustrates how issue-based publics can maintain sustained engagement around structural concerns even when broader discourse trends shift. Importantly, this cluster reflected a high concentration of verified health equity advocates, academic institutions, and non-profit organizations, underscoring the value of trust-building and collective advocacy in marginalized communities.

\subsubsection{Cluster 1 and the Evolving Needs of Marginalized Publics} \label{subsec:marginalized_pub}
Cluster 1 stood out in its consistency and cohesion across the timeline. While discourse around race declined in the broader CDC public early in the pandemic, Cluster 1 remained focused on racial disparities in healthcare, particularly as they intersected with COVID-19 outcomes. Our time-series analysis revealed flashpoints of intensified engagement tied to external socio-political events, including the George Floyd protests and CDC announcements about racism as a public health threat.

The thematic discourse within Cluster 1 moved from highlighting disproportionate mortality and hospitalization rates among BIPOC populations to demanding data justice and structural accountability. These shifts illustrate a critical feature of crisis communication: the evolution of public concerns over time. Cluster 1 encapsulates how marginalized publics respond to both the pandemic and the ways institutions manage it. Their discourse challenges symbolic gestures (e.g., declarations without follow-through) and calls for tangible structural reform. As such, Cluster 1 offers a model for understanding the layered, evolving needs of issue-specific publics, especially those historically underrepresented in public health dialogue.

However, while Cluster 1 continued its engagement with other CDC publics, and was overall one of the more retweeted clusters, it had little overall effect as the decline in overall focus on health inequity at the intersection of race and class. This finding expands on ealier work by Ahmed who found that certain bodies and their stories are made illegible in public discourse due to emotional economies in which certain narratives are more easily absorbed or circulated than others \cite{ahmed2013cultural}. In particular, affectively charged narratives that align with dominant sensibilities—such as hope, resilience, or redemption—are often privileged, while those that center on structural critique, pain, or systemic injustice are marginalized or dismissed. As a result, even when marginalized publics gain visibility through mechanisms like retweets or temporary virality, their ability to reshape dominant narratives remains constrained. This underscores the limits of visibility as a proxy for influence and highlights how affective economies continue to regulate which stories matter and which are rendered unintelligible. In the next sections, we present design and policy recommendations to address this issue.

\subsection{Design Recommendation: Multi-Agent CDC AI Assistants for Diverse Publics} \label{sec:engage_gen_ai}
Building on our analysis, we recommend the development of a multi-agent generative AI system tailored to specific issue publics identified in the CDC Twitter discourse. These AI agents, modeled on clusters such as racial health equity (Cluster 1), school reopening (Cluster 15), and mask guideline clarity (Cluster 7), can deliver targeted, responsive messaging and facilitate two-way communication.

A racial health equity-focused agent, for instance, could proactively disseminate updates on vaccine equity, respond to queries about structural racism in healthcare, and amplify voices from trusted organizations. It can also provide context-aware recommendations (e.g., suggest following key equity advocates), improving both trust and reach. The CDC AI Assistant should operate with transparency, explaining the rationale behind each interaction  \cite{Lubos_LLM_Rec, DiPalma_23}. A similar structure was suggested to suggest new content and social connections in foster care online communities \cite{ammari_ahn_et_al_25}.

Importantly, such systems should be co-designed with public health experts and community members to ensure cultural competence, appropriateness, and ethical data use. This approach aligns with inclusive design principles and expands the CDC's ability to meet different publics "where they are" \cite{mathews2020meet}.

\subsection{Policy Recommendation: Planning for the Long-Now of Crisis Communication} \label{sec:policy_rec}
Traditional crisis communication frameworks often prioritize immediate response. However, our study affirms the need for sustained, adaptive engagement over the long arc of a public health crisis. The CERC model \cite{reynolds2005crisis} emphasizes pre-crisis relationship-building with stakeholders. Our findings extend this principle by highlighting the importance of long-term monitoring and interaction with issue-based publics across platforms.

To operationalize this, the CDC and similar institutions should incorporate real-time modeling of regional discourse dynamics, aligning with work by Wen et al. \cite{wen2022inferring} and Whata \& Chimedza \cite{whata2021machine}. These models can help detect emergent needs, track sentiment changes, and inform timely policy adjustments. Collaborating with local mutual aid networks and community leaders can further bridge the gap between national messaging and local realities \cite{soden_owen_2021, randazzo2025werelosingneighborhoodswere}.

Finally, public health institutions should prepare for the integration of emerging technologies into their communication infrastructure. This includes not only leveraging generative AI tools but also instituting safeguards against their misuse. As our revised public policy discourse framing framework (Figure \ref{fig:park_lee}) suggests, platform affordances, user behaviors, and algorithmic governance must be central to designing resilient communication ecosystems for the crises of tomorrow.

\section{limitations and future work}
This work suggests that crisis organizations like the CDC need to continuously engage with, and follow the discourse directed by different affective \cite{papacharissi2012affective} and issue publics \cite{kim2009issue}, and especially for marginal publics \cite{squires2002rethinking} which might be focused on advocacy for marginalized populations. Future work should engage with different publics to co-design both the policy and design requirements of public policy discourse \cite{park2020stakeholder}. Similar methods have been adopted by Tsai et al. \cite{tsai_et_al_24} when the authors co-designed disaster management systems with indiginous populations. Similarly, AI assistants, like the ones presented in our design section need to be co-designed with the members of different publics, both online and offline, given that both modalities complement each other in crisis management communication \cite{randazzo2025werelosingneighborhoodswere}.

\section{Conclusion}

This study examined the evolving nature of crisis communication within a networked public, focusing on how marginalized communities—particularly those concerned with racial health disparities—participated in and shaped online discourse during the COVID-19 pandemic. We found that although platforms like Twitter allow marginalized publics to maintain visibility, their ability to influence dominant narratives is structurally limited. Cluster 1, composed largely of racial equity advocates and institutions, sustained focus on issues like vaccine access, systemic racism, and data justice throughout the crisis, even as broader public interest declined. These findings affirm the need for inclusive, adaptive, and sustained communication strategies that can meet evolving community needs across different crisis phases. We recommend co-designing AI-enabled tools tailored to issue-specific publics, as well as adopting long-term engagement frameworks that acknowledge and respond to the differentiated temporal trajectories of marginalized communities.


\bibliographystyle{ACM-Reference-Format}
\bibliography{99_refs}

\end{document}